\documentclass[prl,twocolumn,aps, secnumarabic,balancelastpage,amsmath,amssymb,floatfix]{revtex4-1}
\usepackage{graphicx}      
\usepackage{bm}
\usepackage{float}
\usepackage{todonotes}

\newcommand{\eff}{\text{eff}}

\begin{document}
\title{Exponentially Slow Heating in Short and Long-range Interacting Floquet Systems}
\author{Francisco Machado$^{1}$, Gregory D. Meyer$^{1}$, Dominic V. Else$^2$, Chetan Nayak$^{2,3}$, Norman Y. Yao$^1$}
\affiliation{$^{1}$Department of Physics, University of California, Berkeley, CA 97420, USA\\
  $^{2}$Physics Department, University of California, Santa Barbara, CA 93106 USA\\
  $^{3}$Station Q, Microsoft Research, Santa Barbara, CA 93106-6105, USA}

\begin{abstract}

We analyze the dynamics of periodically-driven (Floquet) Hamiltonians with short- and long-range interactions, finding clear evidence for a thermalization time, $\tau^*$, that increases exponentially
with the drive frequency.
We observe this behavior, both in systems with short-ranged interactions,
where our results are consistent with rigorous bounds,
and in systems with long-range interactions, where such bounds do not exist at present.
Using a combination of heating and entanglement dynamics, we explicitly extract the effective energy scale controlling the rate of thermalization. 
Finally, we demonstrate that for times shorter than $\tau^*$, the dynamics of the system
is well-approximated by evolution under a time-independent Hamiltonian $D_\eff$, for both short- and long-range interacting systems. 
\end{abstract}

\maketitle


Periodic driving is a ubiquitous tool
for the controlled manipulation of quantum
systems. Classic examples abound in the context of magnetic resonance spectroscopy,
where a broad class of dynamical decoupling pulse sequences have been developed
to suppress unwanted interactions, both within a system's own degrees of freedom,
as well as with an external environment \cite{slichter2013principles,hahn1950spin,waugh1968approach,rhim1971time,rhim1973analysis,viola1999dynamical,vandersypen2005nmr}. 
Periodic driving has also become a staple in the engineering toolshed of both condensed matter and atomic physics, enabling the realization of topological insulators from nominally trivial band structures \cite{Inoue10,Lindner11a,WangY13,Jiang11b,Thakurathi13} and the generation of synthetic gauge fields for neutral atoms \cite{jaksch2003creation,galitski2013spin,goldman2014light}.

When a generic  system with many degrees of freedom is periodically driven, it typically
absorbs energy from the driving field and heats up to an infinite temperature state \cite{prosen1998time,prosen1999ergodic,d2013many,lazarides2014equilibrium, rigol14,bukov2015universal,ponte2015periodically, Bukov2015,weidinger2017floquet, Luitz2017}, a process
called thermalization \footnote{The analogous phenomenon in undriven systems
is the evolution of a generic state into a thermal state.}. However, when the driving frequency
is high, the Floquet system can only absorb energy from the drive by creating multiple local excitations --- an inefficient process that results in an extremely long thermalization time \cite{abanin2015exponentially,abanin2015rigorous,else2017prethermal,abanin2017effective}.
The system does \emph{eventually} thermalize, but during the time interval
before this occurs, it settles into a ``prethermal'' state \cite{berges2004prethermalization,moeckel2008interaction,gring2012relaxation,marcuzzi2013prethermalization,essler2014quench} that exhibits the hallmarks of
thermal \emph{equilibrium}, albeit at a lower entropy than the true infinite temperature thermal state (which is
only reached at very late times).
In this paper, we characterize and elucidate the mechanism of Floquet thermalization.

\begin{figure}
  \centering
  \includegraphics[width = 3.2in]{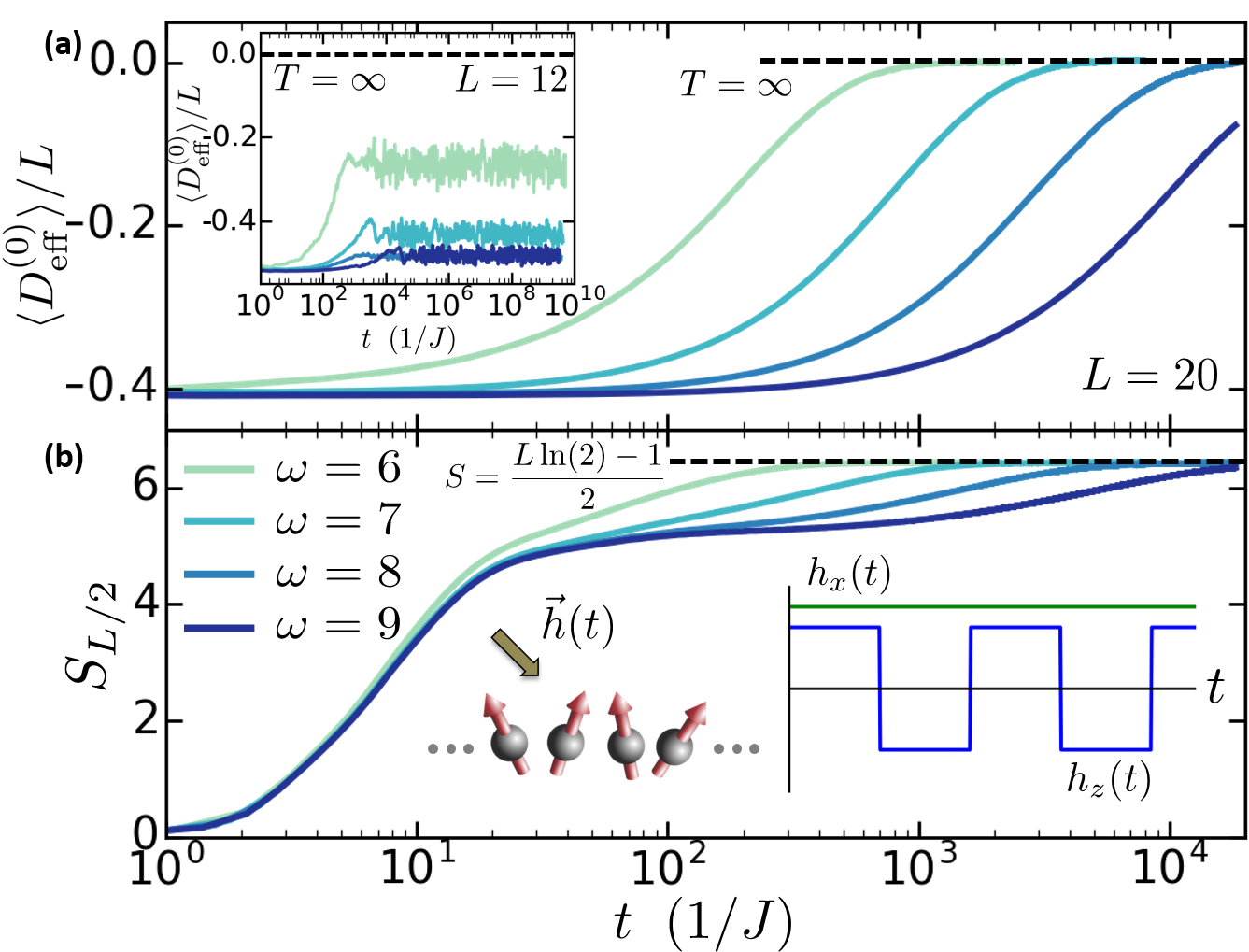}
  \caption{Floquet thermalization dynamics of a long-range interacting spin model with $L=20$.  (a) As the driving frequency is increased, one observes an \emph{exponential enhancement} in the time scale at which the system approaches  infinite-temperature as diagnosed by the energy density, $\langle D_\eff^{(0)}\rangle/L \to 0$. (inset) For smaller system sizes, full thermalization to infinite temperature is never observed even at late times.
  (b) The same exponentially slow thermalization is seen in the time scale where the  half-chain entanglement entropy reaches its infinite temperature value, $\frac{L}{2} \log(2) -0.5$. (inset) Each spin is periodically driven by a time-dependent magnetic field which exhibits a square pulse shape.}
  \label{fig1}
\end{figure}

\begin{figure*}
  \centering
  \includegraphics[width = 7in]{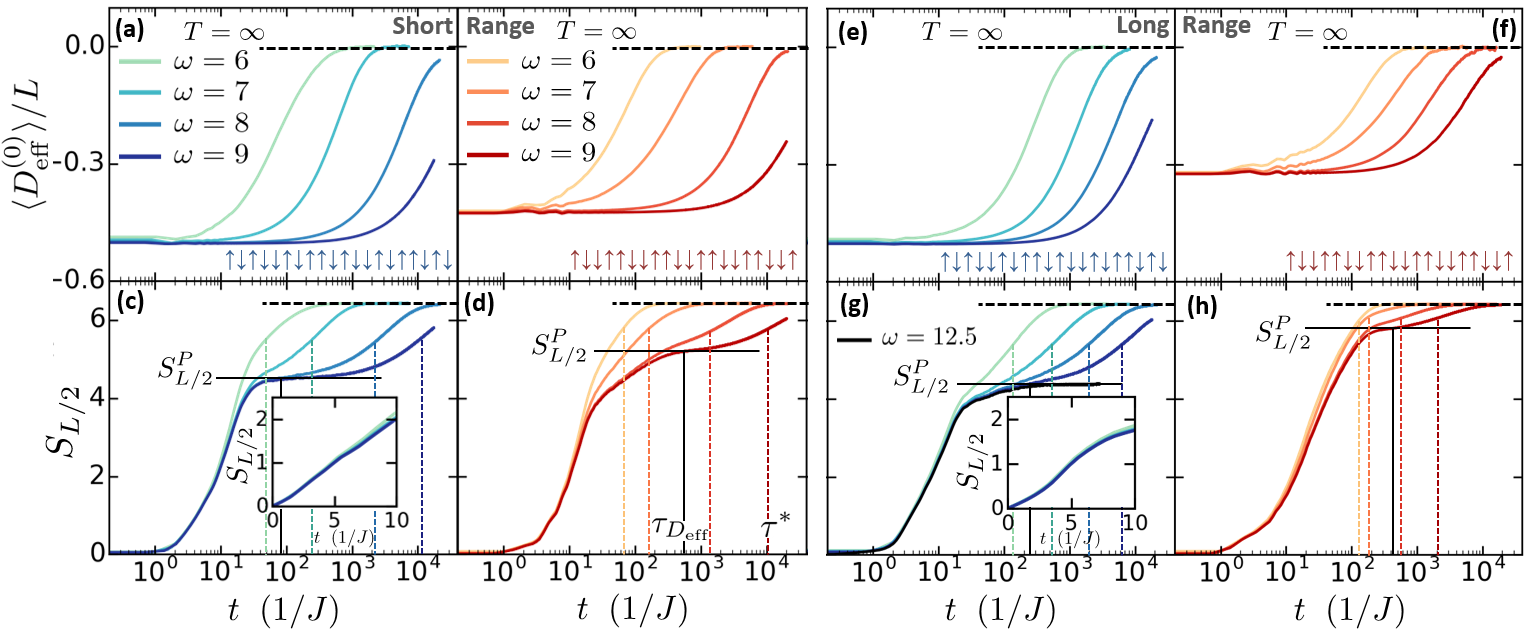}
  \caption{Floquet evolution of both  short- and long-range interacting systems with $L=20$.  a,b) [e,f)] Energy density as a function of time for short-range [long-range] interactions, as measured with respect to the prethermal Hamiltonian $D_{\text{eff}}^{(0)}$ for a low temperature (a[e]) and a high temperature (b[f]) initial state. As one increases the frequency of the periodic drive, one observes an exponential increase in the thermalization time (to infinite temperature). c,d) [g,h)] The half-chain entanglement entropy as a function of time for short-range [long-range] interactions. Two distinct timescales emerge: $\tau_{D_{\eff}}$ and $\tau^*$.  $\tau^*$ corresponds to the thermalization time and is estimated via the colored, dashed vertical lines. $\tau_{D_{\eff}}$ corresponds to the time-scale where the system reaches the prethermal Gibbs state (with entropy $S^P_{L/2}$) of the effective Hamiltonian $D_\eff$, and is indicated via a solid, black vertical line. 
(inset, c)  The initial evolution of $S_{L/2}$ is consistent with a linear light cone in the case of the short-range interactions and deviates from linear growth in the long-range case (inset, g) \cite{hastings2006spectral,hauke2013spread,eisert2013breakdown,gong2014persistence,foss2015nearly,Matsuta2017}.}
  \label{fig2}
\end{figure*}

Using massively parallel  Krylov subspace methods, we explore
the late time dynamics of periodically-driven spin chains with
both short- and long-ranged interactions. 
For short-range interactions and bounded
local Hilbert spaces, seminal recent results have proven that the thermalization time, $\tau^*$, 
increases at least exponentially (up to log corrections) with the frequency of the drive \cite{abanin2015exponentially,abanin2015rigorous,else2017prethermal}.~We provide  the first concrete demonstration of this.
To this end, our results are consistent with those of \cite{bukov2016heating}, which also observed slow heating; but additionally, by directly observing the \emph{exponential scaling} of the thermalization time, we can extract the effective energy scale controlling the Floquet heating rate. 
This is enabled by going to sufficiently large system sizes such that there is a clear separation of scales 
between the local bandwidth and the global many-body bandwidth \footnote{The global
bandwidth is infinite in the thermodynamic limit, but finite in a finite system size.}; indeed, for driving frequencies 
above  the many-body bandwidth, the system is trivially blocked from heating up to infinite temperature (inset, Fig.~\ref{fig1}a).
Moreover, we demonstrate that the half-chain entanglement entropy, $S_{L/2}(t)$,
quickly reaches a plateau value consistent with a prethermal state before saturating to its infinite-temperature value at exponentially-late times \cite{abanin2015exponentially,abanin2015rigorous,else2017prethermal}. On this prethermal plateau, there is an emergent time-independent Hamiltonian, $D_\eff$, that is conserved and generates the time evolution of the system at stroboscopic times $t=mT$ (where $T$ is the period of the drive).

Finally, we also observe exponentially-long thermalization time scales (as well as an emergent $D_\eff$) in a long-range, power-law interacting system, for which no bounds exist in the previous literature; such a result is particularly relevant to isolated quantum optical systems of atoms, ions and molecules, where strong interactions often take the form of long-range coulomb, dipolar, or van der Waals couplings \cite{yan2013observation,zeiher2017coherent,dutt2007quantum,schneider2012experimental}.

\emph{Model and Probes}---We analyze one-dimensional
spin chains whose Floquet evolution is governed by a Hamiltonian with power-law interactions:
\begin{equation}
\label{eq:model}
 {H_\ell}(t) = J \sum\limits_{i<j}\frac{\sigma_i^z \sigma_j^z}{|i-j|^\alpha}  + \vec{h}(t)\cdot\left[\sum\limits_{i} \vec{\sigma_i}\right]+J_x\sum\limits_{\langle i, j\rangle }\sigma^x_i\sigma^x_j \\
\end{equation}
where $ \vec{h}(t) = h_x\hat{x} + (h_y\hat{y} + h_z\hat{z})(1-2\theta(t-T/2))$ (inset, Fig.~\ref{fig1}{b}) \footnote{The parameters used for the remainder of the manuscript are: $J=1$, $J_x = 0.19$, $h_x = 0.21$, $h_y=0.17$, $h_z = 0.13$, $\alpha = 1.25$.}, $\sigma_i^\gamma$ are Pauli operators,  $\omega = 2\pi/T$ is the driving frequency  \cite{abanin2015exponentially,abanin2015rigorous,else2017prethermal}. 
All energies are measured in units in which $J = 1$. We will also consider a short-range interacting model, ${H_s}(t)$, realized by truncating the Ising interaction in $H_\ell$ to nearest and next-nearest neighbor.  


To characterize the Floquet thermalization dynamics, we will begin with two diagnostics (Fig.~1). First, we will use the increase of the energy  averaged over a period of the drive:
$D_\eff^{(0)}   \equiv \frac{1}{T}\int_0^T dt \ {H_l}(t)
=  J \sum\limits_{i<j}\frac{\sigma_z^i \sigma_z^j}{|i-j|^\alpha}  + h_x \sum\limits_{i}  \sigma_x+J_x\sum\limits_{\langle i, j\rangle }\sigma_x^i\sigma_x^{j}$ \footnote{In the short-range case, $D_\eff^{(0)}$ will of
course only contain the nearest and next nearest neighbor Ising terms.}; we note that  $D_\eff^{(0)}$ is actually the first term in an expansion for the prethermal Hamiltonian,
$D_\eff = D_\eff^{(0)} + D_\eff^{(1)}/\omega + D_\eff^{(2)}/\omega^2 + \cdots$, which contains a finite but exponentially-large number of terms \cite{abanin2015exponentially,abanin2015rigorous}. 
To set notation, let us also define  $\mathcal{D}_{\eff}^n$ as the truncation of $D_\eff$ to $n$-th order in $1/\omega$.
As a second diagnostic,  we will investigate the growth of the half-chain entanglement entropy as a function of time:
$S_{L/2} \equiv \text{Tr}(-\rho_{L/2} \ln \rho_{L/2})$ where
$\rho_{L/2}\equiv \text{Tr}_{1\leq i \leq L/2}(|\psi(t)\rangle\langle\psi(t)|)$. 

\emph{Exponentially slow thermalization}---We directly compute the Floquet evolution of up to $L=22$ spins using massively parallel
Krylov subspace techniques \cite{slepc1, slepc2, petsc1}. We consider initial product states with spins
polarized along $\hat{z}$ and control the energy density of the initial state by varying
the number of equally-spaced domain walls that are present. 
We begin with the short-range model, $H_s(t)$  and compute the time evolution of $\langle D_\eff^{(0)}(t)\rangle/L$ for $L=20$ spins at a
variety of driving frequencies (significantly larger than the local energy
scales of the Hamiltonian but smaller than the global many-body bandwidth \footnote{The number of foldings of the quasi-energy spectrum, for $L=20$ at $\omega=8$ is $\sim 5$.}). 
Unlike the small size ($L=12$) exact diagonalization (ED) results (inset, Fig.~\ref{fig1}a), one observes a clear approach to infinite temperature
($\langle D_\eff^{(0)}\rangle/L \to 0$) at late times, as shown in Fig.~\ref{fig2}a.
We define the thermalization time $\tau^*$ as the time at which the energy density
is halfway from its initial value to its infinite temperature value, so that $\tau^*$ is defined by,
$\langle D_\eff^{(0)}(\tau^*)\rangle = 0.5 \langle D_\eff^{(0)}(t=0)\rangle$.
For both low (Fig.~\ref{fig2}a) and high temperature (Fig.~\ref{fig2}b) initial states,
one observes an exponential enhancement of $\tau^*$ as a function of
increasing driving frequency.

\begin{figure}
  \centering
  \includegraphics[width = 3.4in]{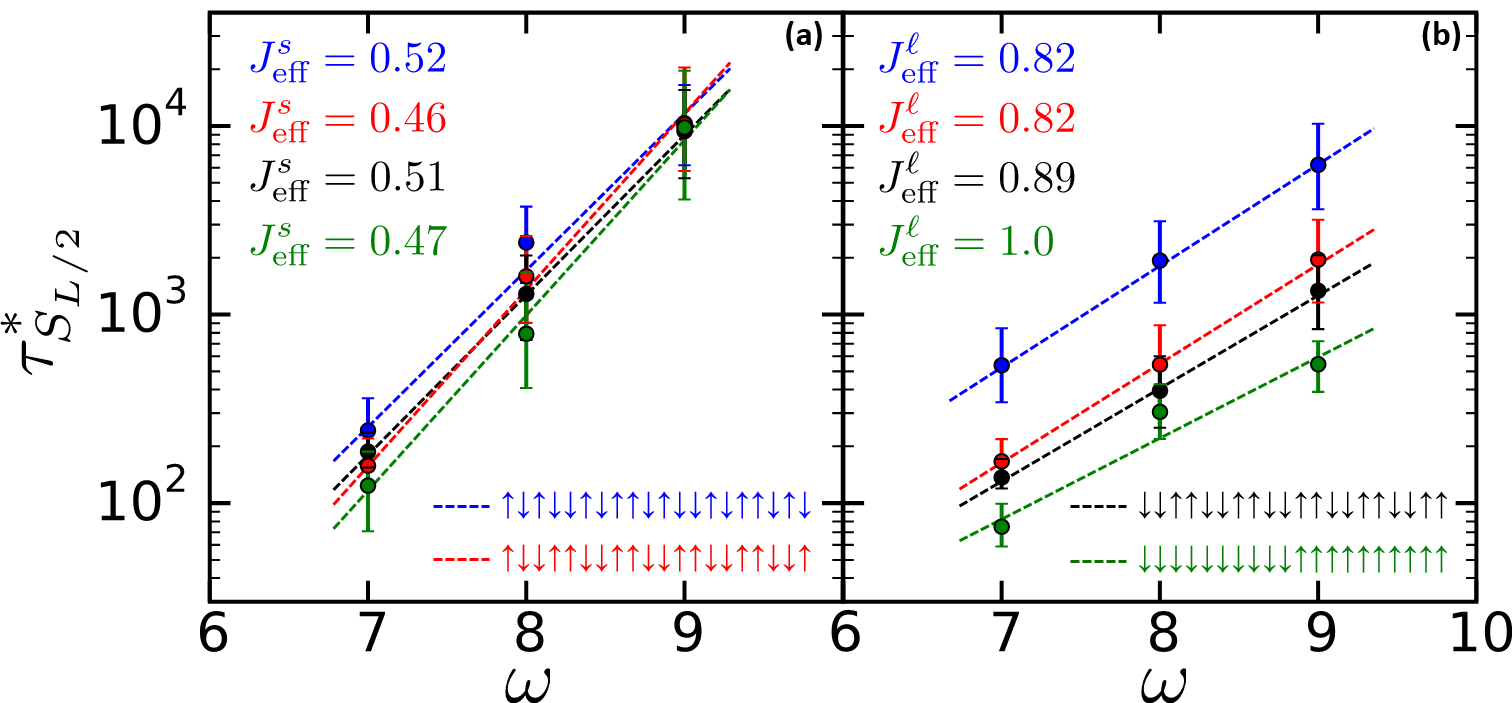}
  \caption{The thermalization time $\tau^*$, as extracted from $S_{L/2}$, as a function of driving frequency for both (a) short- and (b) long-range interactions. The slope provides a direct estimate of $J_{\eff}$, the energy scale, controlling the slow thermalization dynamics. The extracted $J_{\eff}$ is largely independent of initial state (different colors) and is  consistent with its interpretation as an effective local energy scale of the system.  Initial states near the edge of the spectrum exhibit slightly larger $\tau^*$, which can be qualitatively attributed to a reduction of the density of states at these energies. }
  \label{fig3}
\end{figure}

To further probe the exponentially slow heating of the system, we investigate the growth of the half-chain entanglement entropy as a function of time. We expect the evolution of $S_{L/2}(t)$ to be characterized by three distinct regimes: an initial growth period beginning from $S_{L/2}(0)=0$; an intermediate plateau where the entropy reaches its prethermal value, $S_{L/2}^{P}$; and a final plateau once the system has fully thermalized to infinite temperature, with $S_{L/2} = (L\ln(2) - 1)/2$ \cite{page1993average}.
This is indeed born out by the numerics (Fig.~\ref{fig2}c,d).
The time scale at which the entropy is halfway from its prethermal plateau value to its infinite temperature value gives us an alternate definition of $\tau^*$,
$S_{L/2}(\tau^*) = S_{L/2}^{P} + [(L\ln(2) - 1)/2 - S_{L/2}^{P}]/2$,
and has the virtue of not relying upon a choice of operator (such as $\mathcal{D}^n_\eff$) used to probe the state of the system. For both low (Fig.~\ref{fig2}c) and high (Fig.~\ref{fig2}d) temperature initial states,
one observes an exponentially-long heating time scale consistent with that extracted
from $\langle D_\eff^{(0)}\rangle/L$. To this end, Fig.~\ref{fig3}a shows just how well $\tau^*$ fits an
exponential dependence for a variety of different initial states.

There is a second time scale in the problem; namely, the  time, $\tau_{D_\eff}$,  at which
the entanglement entropy reaches its prethermal plateau value, $S_{L/2}^{P}$, as depicted in Fig.~\ref{fig2}c,d. This is the time at which the system \emph{globally} establishes the prethermal
equilibrium-like Gibbs state of $D_\eff$ and is expected to be greater than the \emph{local} thermalization time of $D_\eff$ by a factor of order $\sim L$.
The value of the plateau entropy, $S_{L/2}^{P}$, depends on the inverse temperature of the prethermal ensemble, $\beta_\eff$, which in turn can be directly estimated
using the energy density, $\epsilon$, of the initial state:
$ \epsilon L =
\langle D_{\eff}(t=0)\rangle \approx \text{Tr}\left[D_\eff \ e^{-\beta_\eff D_\eff}\right]$.


To quantitatively verify this relationship, we utilize small size exact diagonalization results on $\mathcal{D}_\eff^{4}$ in order to estimate the entanglement entropy as a function of inverse temperature \cite{suppinfo}. 
In the case of short-range interactions, this approach predicts $S^P_{L/2}= 4.6 \pm 0.4$  and $S^P_{L/2}= 5.4 \pm 0.5$ for low and high temperature initial states, respectively, both in excellent agreement with the numerically observed plateau (Fig.~2c,d).
For long-range interactions, we find that finite size effects in the ED prevent an accurate extrapolation of the entropy and lead to  systematic overestimate of the plateau entropy \cite{suppinfo}.


We now turn to the long-range interacting model, ${H_\ell}(t)$, with power-law $\alpha = 1.25$,  where we again compute $\langle D_\eff^{(0)}(t)\rangle/L$ and $S_{L/2}(t)$. We note that the recent proofs  \cite{abanin2015exponentially,abanin2015rigorous,else2017prethermal} of exponentially-slow heating in Floquet systems seem to be naturally extendable to the case of long-range few-body interactions,
such as the power-law two-body interactions in Eqn.~(\ref{eq:model}).
The intuition is that the system still needs to make many rearrangements, each with a few-body (albeit long-ranged) interaction, in order to absorb energy $\omega$ from the drive. 
Indeed, for both low (Fig.~\ref{fig2}e,g) and high (Fig.~\ref{fig2}f,h) temperature initial states,
we observe exponentially slow heating times as a function of frequency, analogous to  the short-range case.

A few remarks are in order. First, we note that in the long-range model, the early-time
entanglement entropy exhibits a more complex light cone, deviating from the linear one
observed in the short-ranged case (inset, Fig.~\ref{fig2}c,g). 
%
Second, for the same frequencies at which there is already a clear plateau in the short-ranged model, the long-range system exhibits a shoulder with a weak up-slope, which only flattens into a true plateau for larger frequencies (Fig.~2g).
Third, while both the short- and long-range systems exhibit exponentially slow thermalization, there is a clear quantitative difference between the heating rates in the two cases. 

\begin{figure}
  \centering
  \includegraphics[width = 3.0in]{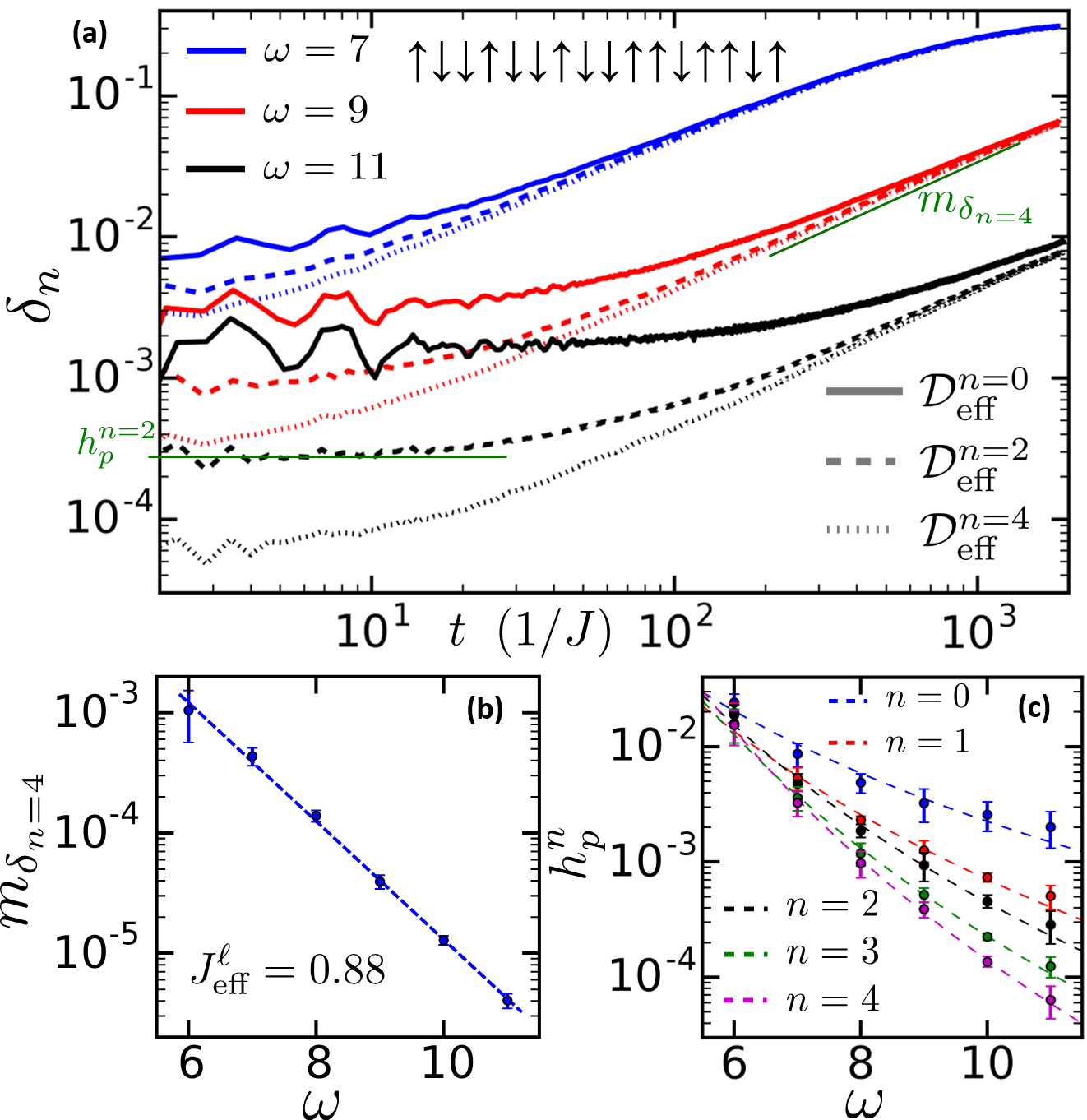}
  \caption{a) The difference, $\delta_n$, of the expectation value of $\mathcal{O} = D_\eff^{(0)}/L$ as a function of time, for a chain of length $L=16$, with different frequencies (colors) and different Magnus truncation orders (line style). The distinct regimes are seen: an initial plateau at short times and a linear increase at late times.  b) Extracted slope of the late time linear regime of $\delta_n$ as a function of frequency. This provides an independent estimate for $J_\eff^\ell$ which is in excellent agreement with that calculated from $\tau^*_{S_{L/2}}$.  c) Plateau height $h_{pl}^{(n)}$ for different Magnus truncation orders, $n$, as a function of frequency. The results are consistent with an $n$-dependent power law.}
  \label{fig4}
\end{figure}

To further explore this, we directly extract the energy scale controlling the exponentially slow heating (i.e.~the effective local bandwidth), 
by fitting $\tau^*$ (extracted from $S_{L/2}$) to $\tau^* \sim e^{\omega/J_\eff}$,
as depicted in Fig.~\ref{fig3}.
In the case of short-range interactions, both low and high temperature initial states give $J_\eff^s \approx 0.5 \pm 0.1$, consistent with the strength of terms in $H_s(t)$. For the long-range interacting model, one finds a larger $J_\eff^\ell \approx 0.9 \pm 0.1$.
Intriguingly, these heating rates yield a ratio, $J_\eff^\ell / J_\eff^s \approx 1.8 \pm 0.2$,  which is  consistent with 
 the ratio of the average strength of all interactions emanating from each site, 
$\left[\sum|i-j|^{-1.25}\right]/\left[1+ 2^{-1.25}\right] \approx 1.6$ \footnote{The analytic ratio averages over all sites of a $L=20$ open chain.}.
We note that the prefactor of the exponential in $\tau^*$ is larger for initial states near
the edges of the spectrum, which could arise from the smaller density of states
there (Fig.~\ref{fig3}) \cite{suppinfo}.

\emph{Long-range prethermal effective Hamiltonian}---We now demonstrate that the time-independent
prethermal Hamiltonian $D_\eff$ is indeed the generator of Floquet dynamics at
stroboscopic times up to  $\tau^*$. Here, we will focus on the more surprising long-range case, leaving the short-range case for the supplementary information \cite{suppinfo}. 
Unlike the question of slow heating, a proof of the existence
of a time-independent $D_\eff$ may need to employ Lieb-Robinson bounds
for long-range interactions \cite{hastings2006spectral,hauke2013spread,eisert2013breakdown,gong2014persistence,foss2015nearly,Matsuta2017}, for which the tightest possible bounds may not
yet have been found for $d<\alpha<2d$, where $d$ is the spatial dimension.
As aforementioned, we not only observe the same exponentially-slow approach to the maximum entropy (consistent with $\langle D_\eff^{(0)}(t)\rangle/L$), but also the presence of a prethermal plateau (for both low and high temperature initial states), indicative of the existence of $D_\eff$ even for long-range interacting systems (Fig.~\ref{fig2}g,h)! 

Further evidence for the existence of a time-independent $D_\eff$ comes from
comparing the system's evolution under the full Floquet unitary,
$U_f \equiv e^{-i \int_0^T {H_l}(t) dt}$, to evolutions under truncations
of the Magnus expansion: $D_\eff = D_\eff^{(0)} + D_\eff^{(1)}/\omega + D_\eff^{(2)}/\omega^2 + \cdots$ at leading order ($\mathcal{D}^0_\eff$), at second order ($\mathcal{D}^2_\eff$), and at fourth order ($\mathcal{D}^4_\eff$). 

In Fig.~\ref{fig4}{a}, we plot $\delta_n = |\langle \mathcal{O}\rangle_{U_f} -\langle \mathcal{O}\rangle_{\mathcal{D}_\eff^{n}}|$,
as a function of time for different frequencies and different Magnus truncation orders, with operator $\mathcal{O}=D_\eff^{(0)}/L$
(other choices
of local operators exhibit similar results but this one has the cleanest numerics \cite{suppinfo}). Here, $\langle \mathcal{O} \rangle_{H}$ is the expectation value of $\mathcal{O}$ evolved under $H$; thus, $\delta_n(t)$ captures the time-dependent difference in the expectation value of $\mathcal{O}$ evolved under the full Floquet unitary versus under different approximations to $D_\eff$.
Inspection reveals two essential features: a short-time plateau \footnote{The nature of the short time dynamics is dependent on both the operator considered  and the truncation order of $D_\eff$. Other operators are considered in detail in the supplementary information \cite{suppinfo}. That there is a plateau at short times arises from the close relationship between $D_\eff^{(0)}$ and $\mathcal{D}_\eff^{n}$. As the systems thermalizes with respect to $\mathcal{D}_\eff^{n}$, the expectation value of $D_\eff^{(0)}$ will not change significantly as it approaches its thermal average \unexpanded{$\langle D_\eff^{(0)} \rangle_{\mathcal{D}^n_\eff}= \text{Tr}(D^{(0)}_\eff e^{-\beta_\eff^n \mathcal{D}^n_\eff})/Z $}.
  The value of the plateau then corresponds to the difference in the thermal value of $D_\eff^{(0)}$ calculated with respect to $\mathcal{D}_\eff^n$ and the full $D_\eff$. By varying $n$, one changes both the Hamiltonian to which the system thermalizes as well as the effective temperature of the prethermal regime $\beta_\eff^n$, leading to a non-trivial dependence of the plateau value with both $n$ and $\omega$ (but  expected to monotonically decrease as either increases).} whose value depends on both $n$ and $\omega$, followed by linear growth at late times that seems to converge for the different truncation orders. To understand these features, we note that there are two contributions to $\delta_n(t)$.

First, since $D_\eff$ approximates the full Floquet evolution only up to a time scale $\tau^* \sim e^{\omega/J_\eff}$, one expects the exponentially slow accumulation of errors, $\delta \sim t e^{-\omega/J_\eff}$. Second, even at short times, one expects a finite discrepancy to arise simply from the fact that the $n^{\textrm{th}}$ order Magnus approximation still differs from $D_\eff$ (e.g.~by terms such as $D_\eff^{(n+1)}/\omega^{n+1} +D_\eff^{(n+2)}/\omega^{n+2} + \cdots$). This second point explains the qualitative dependence of the plateau value on $n$ and $\omega$. In particular, larger $n$ and larger $\omega$ both lead to a smaller initial plateau for  $\delta_n(t)$ since they correspond to decreasing the effect of higher-order terms in the expansion; by measuring the plateau height $h_p$ as a function of frequency, we find that it is consistent with $h_p \sim \omega^{-\gamma(n)}$, where $\gamma$ is an $n$-dependent power-law (Fig.~4c).

Finally, the observed linear growth of  $\delta_n(t)$ at late times is consistent with the exponentially slow accumulation of errors, $\delta \sim t e^{-\omega/J_\eff}$, and enables another independent extraction of $J_\eff$. In particular, as shown in Fig.~4b, by plotting the slope of the late time growth of  $\delta_n(t)$ as a function of the frequency, one obtains $J_\eff \approx 0.88\pm0.05$ consistent with that calculated from the entanglement entropy in Fig.~3.

\emph{Conclusion}---Despite their ubiquity, periodically-driven Floquet systems have generally not shown distinct phases of matter.
This is largely due to their tendency to heat up to infinite temperature, except in certain exceptional cases, such as free fermion systems (e.g.~topological insulators
 \cite{Inoue10,Lindner11a,WangY13,Jiang11b,Thakurathi13}), and strongly-disordered one-dimensional (and possibly, two-dimensional)
systems in the many-body localized phase \cite{basko2006metal,gornyi2005interacting,huse2013localization,bahri2013localization,chandran2014many,nandkishore2015many}. 
In the high-frequency limit, however,
we have shown that there is an exponentially-long time interval during which a
system may, as it would in true thermal equilibrium, realize phases of matter
and phase transitions between them, including
certain phases that do not exist in undriven systems \cite{else2017prethermal,zhang2016observation,choi2017observation}.

On the experimental side, prethermalization provides a straightforward technique for extending the thermalization time-scales of Floquet systems. This enables experiments to work in parallel with theory in realizing and studying novel out-of-equilibrium phases. This formalism also enables the engineering of quantum evolution similar to dynamical decoupling and other techniques more common in the magnetic resonance community.

We gratefully acknowledge the insights of and  discussions with E. Altman, B. Bauer, M. Bukov, P. Hess, V. Khemani, C. Monroe, M. Zaletel, J. Zhang.
This work was supported, in part by, the NSF PHY-1654740, the Miller Institute for Basic Research in Science, and the LDRD Program of LBNL under US DOE Contract No.~DE-AC02-05CH11231. D.V.E. is supported by the Microsoft Corporation.

\bibliography{pretherm}

\end{document}


\title{Supplementary Information for Exponentially Slow Heating in Short and Long-range Interacting Floquet Systems}
\author{Francisco Machado$^{1}$, Gregory D. Meyer$^{1}$, Dominic Else$^2$, Chetan Nayak$^{2,3}$, Norman Y. Yao$^1$}
\affiliation{$^{1}$Department of Physics, University of California, Berkeley, CA 97420, USA\\
  $^{2}$Physics Department, University of California, Santa Barbara, CA 93106 USA\\
  $^{3}$Station Q, Microsoft Research, Santa Barbara, CA 93106-6105, USA}

\maketitle

\section{Calculation of $D_\eff$}

In this section we compute the prethermal effective Hamiltonian $D_\eff$ of our periodically driven system. This time-independent Hamiltonian is the approximate generator of stroboscopic time evolution until $\tau^*$. We obtain $D_\eff$ by approximating the time evolution under one period, $U_f$, by a truncated Magnus expansion, leading to a representation of $D_\eff$ as a power series in the period of the drive $T = 2\pi/\omega$.

Consider the time evolution under the Hamiltonian described in \refeq{1} of the main text:
\begin{equation}
  H_\ell(t) =
  \begin{cases}
    \left(J \sum\limits_{i<j}\frac{\sigma^z_i \sigma^z_j}{|i-j|^\alpha} +J_x\sum\limits_{\langle i, j\rangle }\sigma^x_i\sigma^x_j  + \sum\limits_{i} h_x\sigma^x_i\right) + \left(\sum\limits_{i} h_y \sigma^y_i + h_z\sigma^z_i\right) = D + E & \quad \text{for } 0<t<\frac{T}{2}\\
    \left(J \sum\limits_{i<j}\frac{\sigma^z_i \sigma^z_j}{|i-j|^\alpha} +J_x\sum\limits_{\langle i, j\rangle }\sigma^x_i\sigma^x_j  + \sum\limits_{i} h_x\sigma^x_i\right) - \left(\sum\limits_{i} h_y\sigma^y_i + h_z\sigma^z_i \right) = D - E& \quad \text{for } \frac{T}{2}<t<T
  \end{cases},
\end{equation}
where $D$[$E$] is the time [in]dependent component of $H_\ell(t)$. The term $E$ can be thought of as a magnetic field with a square wave time profile in the $\hat{y}$ and $\hat{z}$ directions. As in the main text, we define the analogous short-range model $H_s(t)$ by truncating the Ising interaction to nearest and next nearest neighbor.

The evolution under a period can be succinctly written as:
\begin{equation}
  U_f = \exp\left\{-i\frac{T}{2}(D-E)\right\}\exp\left\{-i\frac{T}{2}(D+E)\right\}.
\end{equation}
$U_f$ can now be cast as the exponential of an effective Hamiltonian:
\begin{align}
  U_f \approx &\exp\{-iT D_\eff\} = \exp\left\{ -i\frac{T}{2} \left(D-E + D+E\right) +\frac{1}{2} \left(-i\frac{T}{2}\right)^2 [D-E,D+E] +...\right\} 
\end{align}
Upon algebraic simplification and collection of terms, $D_\eff$ can be recovered as a sum of products of the terms $D$ and $E$:
\begin{align}\label{eq:Deff}
  D_\eff &= D + \frac{i}{2}\frac{T}{2}(ED-DE) - \frac{1}{6}\left(\frac{T}{2}\right)^2 ( EED -2EDE + DEE)\\\notag
   +& \frac{i}{24}\left(\frac{T}{2}\right)^3 \big[ (EDDD + DEEE -EEED - DDDE) + 3(EEDE + DDED - EDEE - DEDD))\big]\\\notag
   +& \frac{1}{360}\left(\frac{T}{2}\right)^4 \Big[ -27(EDDED + DEDDE) + 23(DDEDE +EDEDD) + 18(EDDDE + EEDEE) + 8DEDED  \\\notag
  &\quad    -12 (EEEDE + EDEEE) -7(EEDDD + DDDEE) + 3(DEEEE + EEEED) - 2(DEEDD + DDEED)\Big]  \\\notag
 +& \quad...\quad\notag
\end{align}
Although cumbersome, this formulation of $D_\eff$ provides a straightforward numerical implementation within the SLEPc and PETSc libraries \cite{slepc1, slepc2, petsc1} as one can obtain all orders of $D_\eff$ in terms of only $D$ and $E$. \req{eq:Deff} holds regardless of the form of its interacting terms, so it applies to both the short- and long-range models. As per the main text, we define $\mathcal{D}_\eff^n$ as the truncation of $D_\eff$ to $n$-th order in $1/\omega$.









\section{$\tau^*$ as a function of initial energy density}

The existence of a prethermal regime has been proven as an upper bound in the difference of time evolved operators under the full Hamiltonian and $D_\eff$ \cite{abanin2015exponentially,abanin2015rigorous,else2017prethermal}. The generality of this approach leaves an open question: what is the effect of different initial states in the thermalization time scale of a system? In this section we attempt to shed some light onto this question by analyzing how the thermalization time scale, $\tau^*$, changes as a function of energy density of the initial state (measured with respect to $ D_\eff^{(0)}$) for both short- and long-range interacting systems.

We estimate $\tau^*$ in two different ways - using the evolution of the entanglement entropy $S_{L/2}(t)$, and of the energy density $\langle \mathcal{D}_\eff^{(2)}\rangle/L$. Firstly, we estimate $\tau^*$ to be the time when the entanglement entropy is half-way between its prethermal plateau $S_{L/2}^P$ and its final value of $(L\ln(2)-1)/2$ \cite{page1993average}:
\begin{equation}
  S_{L/2}(\tau^*) = S_{L/2}^P + \frac{1}{2}\left[ \frac{(L\ln(2)-1)}{2} - S_{L/2}^P\right]\ .
\end{equation}
Given the large size of our system, we are unable to compute directly $S_{L/2}^P$ as a function of the initial energy density (since that would require full exact diagonalization). With this constraint, we instead estimate $S_{L/2}^P$ using the value $S_{L/2}(t)$ when we observe the system has reached a plateau at frequency $\omega=9$:
\begin{equation}\label{eq:tauSL2}
  S_{L/2}^P \approx S_{L/2}(t_{pre})\ ,
\end{equation}
where we have used $t_{pre} = 300,200$ for the short- and long-range models respectively. Secondly, we estimate $\tau^*$ to be the time when the energy is half-way between its initial value and its infinite temperature value $\langle \mathcal{D}_\eff^n(t)\rangle\to 0$:
 \begin{equation}\label{eq:tau_en}
   \langle D_\eff(\tau^*)\rangle = \frac{\langle D_\eff (0)\rangle}{2}\ .
 \end{equation}
 \req{eq:tau_en} contains an ambiguity as to which order one should consider for $D_\eff$. Performing the analysis with different $\mathcal{D}_{\eff}^n$, one observes no qualitative change in the results, so we choose $\mathcal{D}_\eff^2$ for the remainder of this work.

\begin{figure}
  \centering
  \includegraphics[width = 6.0in]{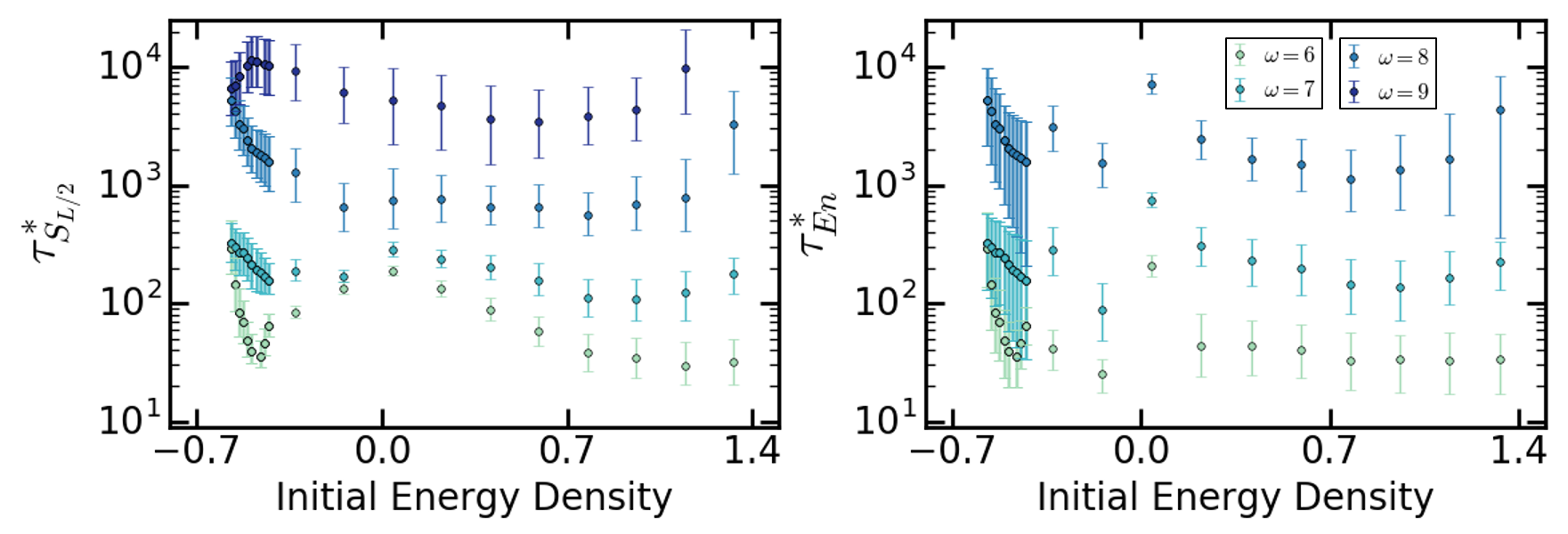}
  \caption{$\tau^*$ of the short-range interacting model as a function of the energy density of the initial state (measured with respect to $D_\eff^{(0)}$) and the frequency of the drive. {\bf a)} $\tau^*$ as estimated using the entanglement entropy of the system. {\bf b)} $\tau^*$ as estimated using the evolution of the energy density. In both cases we observe an overall independence on the initial state except near the center and edges of the spectrum, where we believe our estimation scheme and the change in density of states lead to the deviations, respectively.}
  \label{fig:ShortTau}
\end{figure}

 We now analyze how $\tau^*$ varies for different initial states. We consider initial product states with spins polarized along $\hat{z}$ and control the energy density by varying the number of equally spaced domain walls. 
 In \rfig{fig:ShortTau}, we consider  $\tau^*$ for the short-range interacting system at different frequencies using both entanglement entropy, \rfigP{fig:ShortTau}{a}, and energy density \rfigP{fig:ShortTau}{b}. In both cases we observe the qualitatively similar  behaviors. As a function of  frequency, we observe an exponential dependence across the entire set of initial states, as expected from the state independent results proven in  \cite{abanin2015exponentially,abanin2015rigorous,else2017prethermal}. We also observe no large dependence on the energy density, except near the center and at the edges.
For the former, the closeness of $S_{L/2}^P$ and the initial energy density to their infinite temperature values limits our ability to correctly estimate $\tau^*$.
 For the latter, a lower density of states is expected to decrease the rate at which the system is able to absorb energy from the drive leading to an increase in $\tau^*$, as highlighted in \reffig{3}b of the main text.

\begin{figure}
  \centering
  \includegraphics[width = 6.0in]{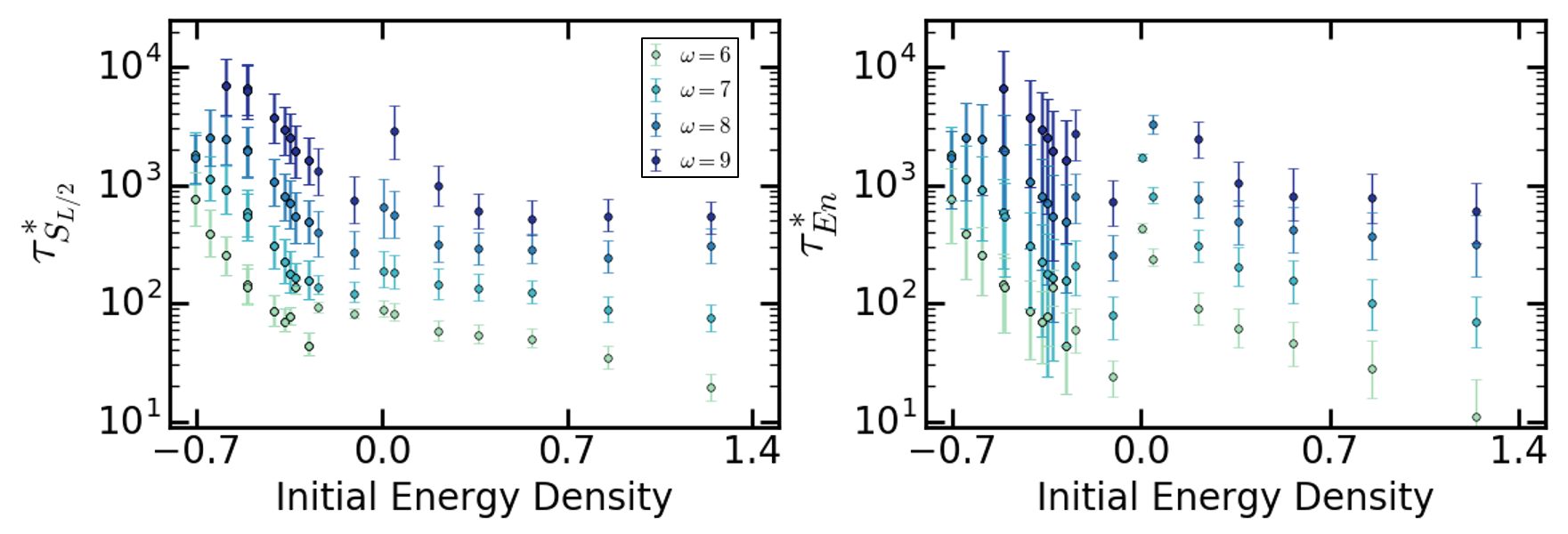}
  \caption{$\tau^*$ of the long-range interacting model as a function of the energy density of the initial state (measured with respect to $D_\eff^{(0)}$) and the frequency of the drive. {\bf a)} $\tau^*$ as estimated using the entanglement entropy of the system $\tau_{S_{L/2}}^*$. {\bf b)} $\tau^*$ as estimated using the evolution of the energy density $\tau_{En}^*$. We observe the same qualitative behavior as in the short-range interacting model, depicted in \rfig{fig:ShortTau}, but with a larger increase of $\tau^*$ at the edge of the spectrum and a smaller dependence with the frequency of the drive, in agreement with the analysis presented in \reffig{3} of the main text.}
  \label{fig:LongTau}
\end{figure}

In \rfig{fig:LongTau} we perform the analogous analysis for the long-range interacting system. Again we observe the same qualitative behavior when estimating $\tau^*$ using the entanglement entropy, \rfig{fig:LongTau}{a}, and the energy density, \rfig{fig:LongTau}{b}. Moreover, both short- and long-range interacting systems present the same overall qualitative features. We note, however, two important differences between the two. In the long-range model there is a more pronounced increase in $\tau^*$ near the edges of the spectrum. This is in agreement with our understanding that this phenomena arises as a density of states effect, since the long-range model has a smaller density of states near the edge of the spectrum. Moreover, the frequency has a smaller impact on $\tau^*$ in the long-range model across the entire spectrum, consistent with the results presented in \reffig{3} for a few different initial states.

\section{Estimating the Entanglement Entropy Plateau Height $S^P_{L/2}$}

\begin{figure}
  \centering
  \includegraphics[width=5.0in]{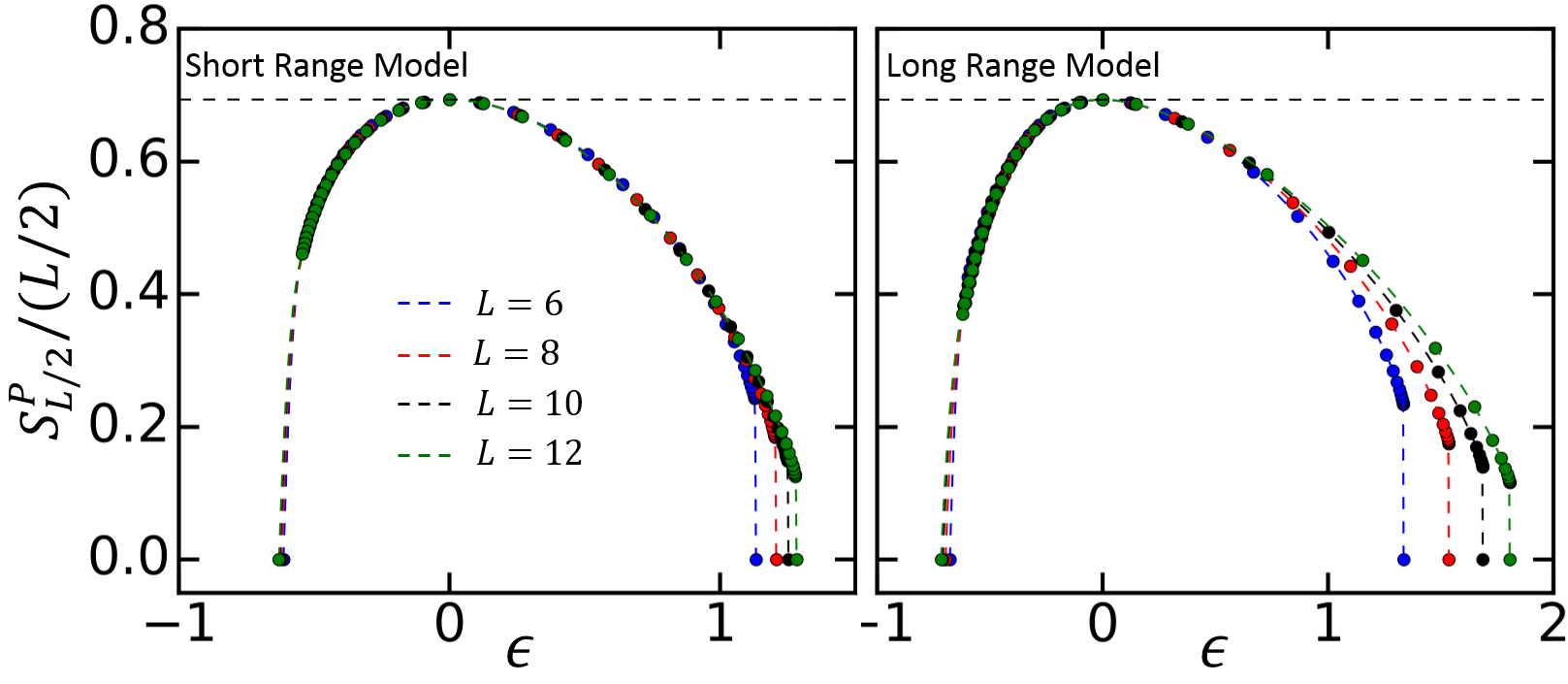}
  \caption{{\bf a[b])} Half-chain entanglement entropy density $S_{L/2}/(L/2)$ as a function of the energy density $\epsilon$ of the thermal state of the short- [long-]range model and for different system sizes $L=6,8,10,12$. Dashed line corresponds to a cubic interpolation to the data points. For negative energy density, $S_{L/2}/(L/2)$ is approximately constant as a function of system size for both short- and long-range models. However, for positive $\epsilon$, there is a large size effect for the long-range model, which limits the validity of our extrapolation.}
\label{fig:EntanglementEntropy}
\end{figure}

In this section, we describe a way to obtain $S_{L/2}^P$ (for $L=20$) by estimating the entanglement entropy of $\mathcal{D}_\eff^n$ on both short- and long-range models  using  exact diagonalization (ED) results from smaller system sizes, $L=6,8,10,12$.
Having obtained the eigenspectrum of $\mathcal{D}^4_\eff$ it becomes straightforward to both compute the energy density $\epsilon$ and $S_{L/2}$ of a thermal state of $D_\eff$ as a function of inverse temperature $\beta$:
\begin{equation}
  L\epsilon(\beta) = \langle \mathcal{D}^4_\eff\rangle_\beta= \frac{\mathrm{tr}( \mathcal{D}^4_\eff e^{-\beta \mathcal{D}^4_\eff})}{Z} \quad \text{ and } \quad S_{L/2}^P(\beta) = \mathrm{Tr}(-\rho_{L/2}^\beta \ln \rho_{L/2}^\beta)
\end{equation}
where $Z = \mathrm{Tr}(e^{-\beta \mathcal{D}^4_\eff})$ and $\rho$ is the reduced density matrix defined as:
\begin{equation}
  \rho_{L/2}^\beta = \mathrm{Tr}_{i\le 1 \le L/2}\left( \frac{e^{-\beta \mathcal{D}^4_\eff}}{Z}\right)
\end{equation}
Having computed both $\epsilon(\beta)$ and $S_{L/2}^P(\beta)$ for different $\beta$, one obtains implicitly the entanglement entropy as a function of the energy density, $S_{L/2}(\epsilon)$, as illustrated in \rfig{fig:EntanglementEntropy}. We have also included the limiting cases of zero and infinite temperature that corresponds to the edge and center of the spectrum ($\epsilon=0$) respectively, where the entanglement entropy is $S_{L/2}^P=0$ and $S_{L/2}^P = \frac{L}{2} \ln(2)$.

From these data, we construct a cubic extrapolation $s^{est}(\epsilon)$ of $S_{L/2}^P/(L/2)$. Given the small finite size effect of $S_{L/2}^P/(L/2)$  for $\epsilon < 0$ we assume that $s^{est}(\epsilon<0)$ remains constant as we vary the system size. For $\epsilon>0$, increasing the system size leads to an increase in the width of the spectrum. As a result, in order to use $S_{L/2}$  at system size $L'=12$ to estimate $S_{L/2}$ at system size $L=20$, we need to rescale the energy density before using our interpolation. The resulting estimate for $S_{L/2}^P$ is given by:
\begin{equation}\label{eq:Sest}
  S_{L/2}^{Est} = \frac{L}{2} \times s^{est}\left( \frac{\epsilon_L}{\epsilon_{L}^{max}} \epsilon_{L'}^{max}\right)\ ,
\end{equation}
where $\epsilon_L$ is the initial energy density in the $L=20$ state, while $\epsilon_{l}^{max}$ is the edge of spectrum at system size $l$. In the case of $l=L$, we consider the energy density of the fully polarized state, while in $l=L'$ it can be obtained exactly from ED. This rescaling serves two purposes: 1) ensuring that the argument falls inside the domain of the interpolation, and 2) attempting to account for  finite size effects. From \rfig{fig:EntanglementEntropy} we observe that a smaller $L$ leads to a larger slope near the edge of the spectrum. This leads to an overestimation of $S_{L/2}^P$ in the long-range case as observed in our results.
Finally, since we are evolving a pure initial state unitarily, the system will never reach the true thermal ensemble, leading to a correction of $0.5$ in \req{eq:Sest} \cite{page1993average}:
\begin{equation}
  S_{L/2}^{Est} =\frac{L}{2} \times s^{est}\left( \frac{\epsilon_L}{\epsilon_{L}^{max}} \epsilon_{L'}^{max}\right) - 0.5
\end{equation}

\section{Estimating Errors in the Numerics}

In order to quantify our uncertainty in the extraction of quantities from our numerics we define the following procedures as our uncertainty.
\begin{itemize}
\item $\tau^*$ - The extraction of $\tau^*$ depends on the estimation of $S_{L/2}^P$. Due to the simple method considered in \req{eq:tauSL2} we consider our uncertainty regime in $\tau^*$ as:
  \begin{equation}
  S_{L/2}(\tau^*_{min}) = S_{L/2}^P + 0.35\times \left[ \frac{(L\ln(2)-1)}{2} - S_{L/2}^P\right]\ ;\; S_{L/2}(\tau^*_{max}) = S_{L/2}^P + 0.65\times \left[ \frac{(L\ln(2)-1)}{2} - S_{L/2}^P\right].
  \end{equation}
  We estimate our error as the maximum deviation between $|\tau^*-\tau^*_{min}|$ and $|\tau^*-\tau_{max}^*|$. In the case of the extration of $\tau^*$ from the energy density we apply the same criterion.
\item $m_{\delta_n}$ - The uncertainty in this quantity arises from the impact of the plateau physics, that can change the fit to a straight line, as well as the choice of the region where we observe the linear regime. We account for these phenomena by dividing the range where we observe the linear effect into six equally sized sub regions. By applying the fit within each of the regions, we obtain an estimate of $m_{\delta_n}$. While the true estimate becomes the average of these values, we take the uncertainty as the $2\sigma$ standard deviation. 
\item $h_p^{n}$ - Similar to $m_{\delta_{n}}$ we define a region where we observe the existence of the plateau. $h_p^n$ is then given by the average of the $\delta_n$ at these points, while we take the uncertainty as the $2\sigma$ standard deviation. 
\end{itemize}

\section{Prethermal effective Hamiltonian for short-range interacting system}

In \reffig{4} of the main text, we analyzed the role of $D_\eff$ as the approximate generator of stroboscopic time evolution for the long-range interacting model. In this section we supplement those results by studying the short-range model using $\mathcal{O}=D_\eff^{(0)}/L$ as well as considering other local operators in both short- and long-range models.
Analogous to the results presented in \reffig{4} of the main text, we consider the difference $\delta_n$ in the expectation value of $\mathcal{O} = D_\eff^{(0)}/L$ when time evolved under $\mathcal{D}_\eff^n$ or the full Floquet unitary in the short-range interacting model with  $L=16$. 

\begin{figure}
  \centering
  \includegraphics[width = 5.0in]{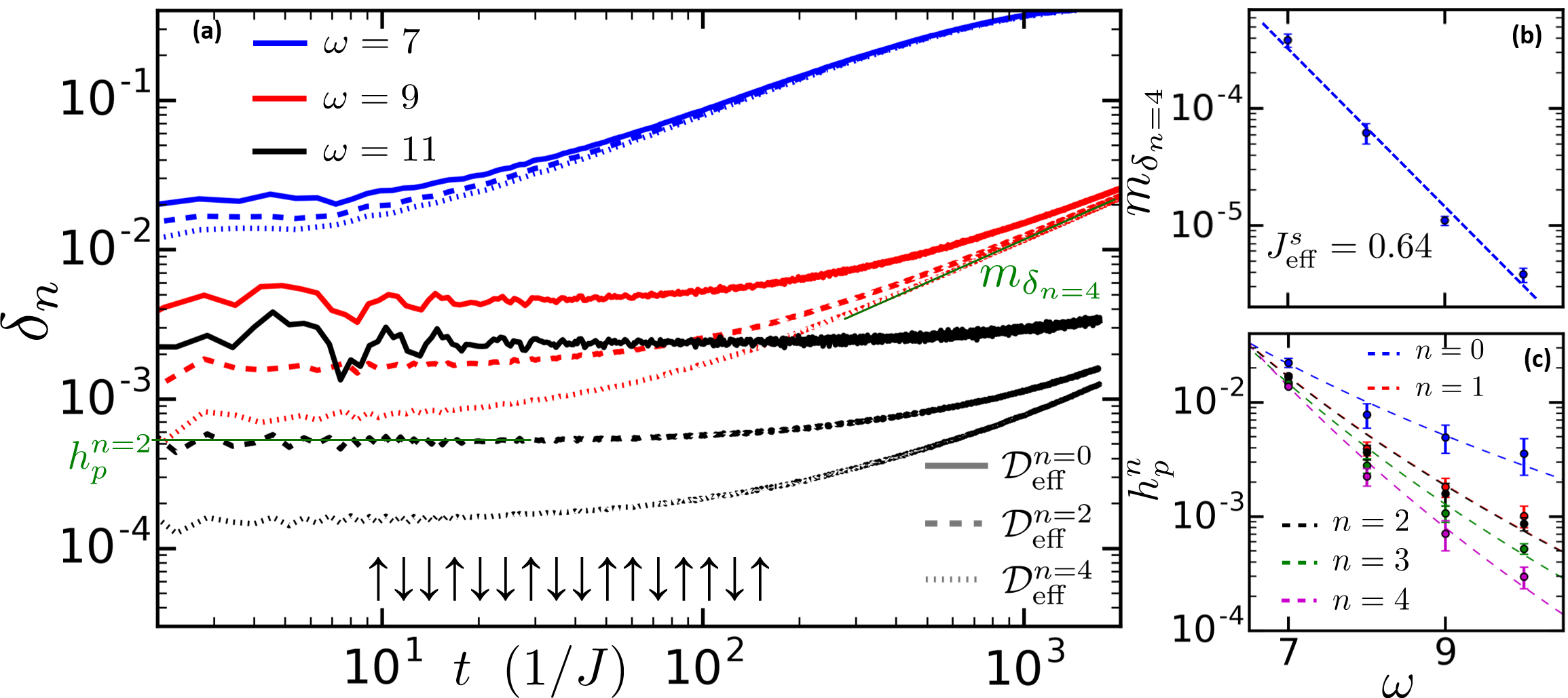}
  \caption{a) The difference, $\delta_n$, of the expectation value of $\mathcal{O} = D_\eff^{(0)}/L$ as a function of time, for a chain of length $L=16$ with short-range interactions, with different frequencies (colors) and different Magnus truncation orders (line style). The distinct regimes are seen: an initial plateau at short times and a linear increase at late times.  b) Extracted slope of the late time linear regime of $\delta_n$ as a function of frequency. This provides an independent estimate for $J_\eff^\ell$ which is in agreement with that calculated from $\tau^*_{S_{L/2}}$, in \reffig{3} of the main text.  c) Plateau height $h_{pl}^{(n)}$ for different Magnus truncation orders, $n$, as a function of frequency. The results are consistent with an $n$-dependent power law.}
  \label{fig4SM}
\end{figure}

In \rfig{fig4SM}a we observe the same qualitative behavior as in the long-range interacting system analyzed in the \reffig{4}a of the main text. In particular we observe the same initial plateau originating from the difference between $\mathcal{D}_\eff^n$ and $D_\eff$ as well as the late time linear regime. Immediately, one notices that for the same range of frequencies, the linear regime of $\delta_n$ occurs at later times corresponding to a slower linear growth. In \rfig{fig4SM}{b,c} we quantify these aspects by analyzing the frequency dependency of both the slope of the linear regime, and the height of the plateaus. From the linear slope, we extract an effective interaction strength $J_\eff^s = 0.6 \pm 0.1$ which is in agreement with the results from the analysis presented in \reffig{3} in the main text, $J_\eff^s=0.5\pm 1$. Regarding the plateau height we observe a similar power law dependence with frequency $h_p \sim \omega^{-\eta}$ as in the long-range case, but with larger values of $\eta$.

\begin{figure}[H]
  \centering
  \includegraphics[width = 5.0in]{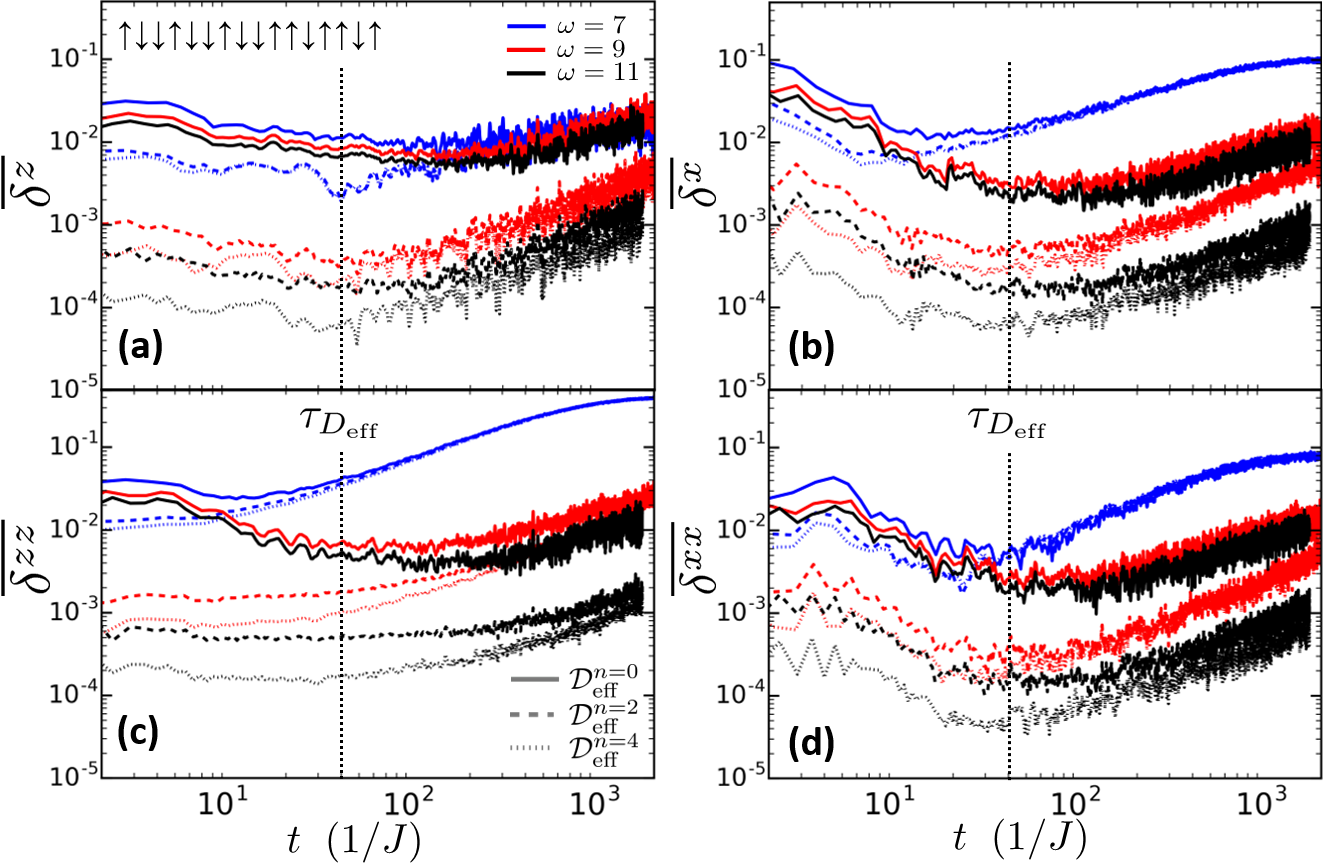}
  \caption{Difference in local operators, ($\sigma_i^z$ [a],$\sigma_i^x$ [b],$\sigma_i^z\sigma_{i+1}^z$ [c],$\sigma_i^x\sigma_{i+1}^x$ [d])  when evolved under the full evolution and $\mathcal{D}_\eff^{n}$ of the short-range model. Similar to \reffig{4} and \rfig{fig4SM} we observe a late time linear regime, corresponding to the linear accumulation of error. At early times we observe a complex behavior arises from the thermalization dynamics to $\mathcal{D}_\eff^n$. For $t>\tau_{D_\eff}$, as extracted from \rfig{Plateau}, the error follows the same behavior as in \reffig{4}a and \rfig{fig4SM}a when $\mathcal{O} = D_\eff^{(0)}/L$. Despite the initial behavior, an increase in $n$ leads to an earlier onset of the linear regime, as expected from $D_\eff$ being the approximate generator of time evolution, consistent with previous theoretical results.}
  \label{LocalShort}
\end{figure}

\begin{figure}[H]
  \centering
  \includegraphics[width = 5.0in]{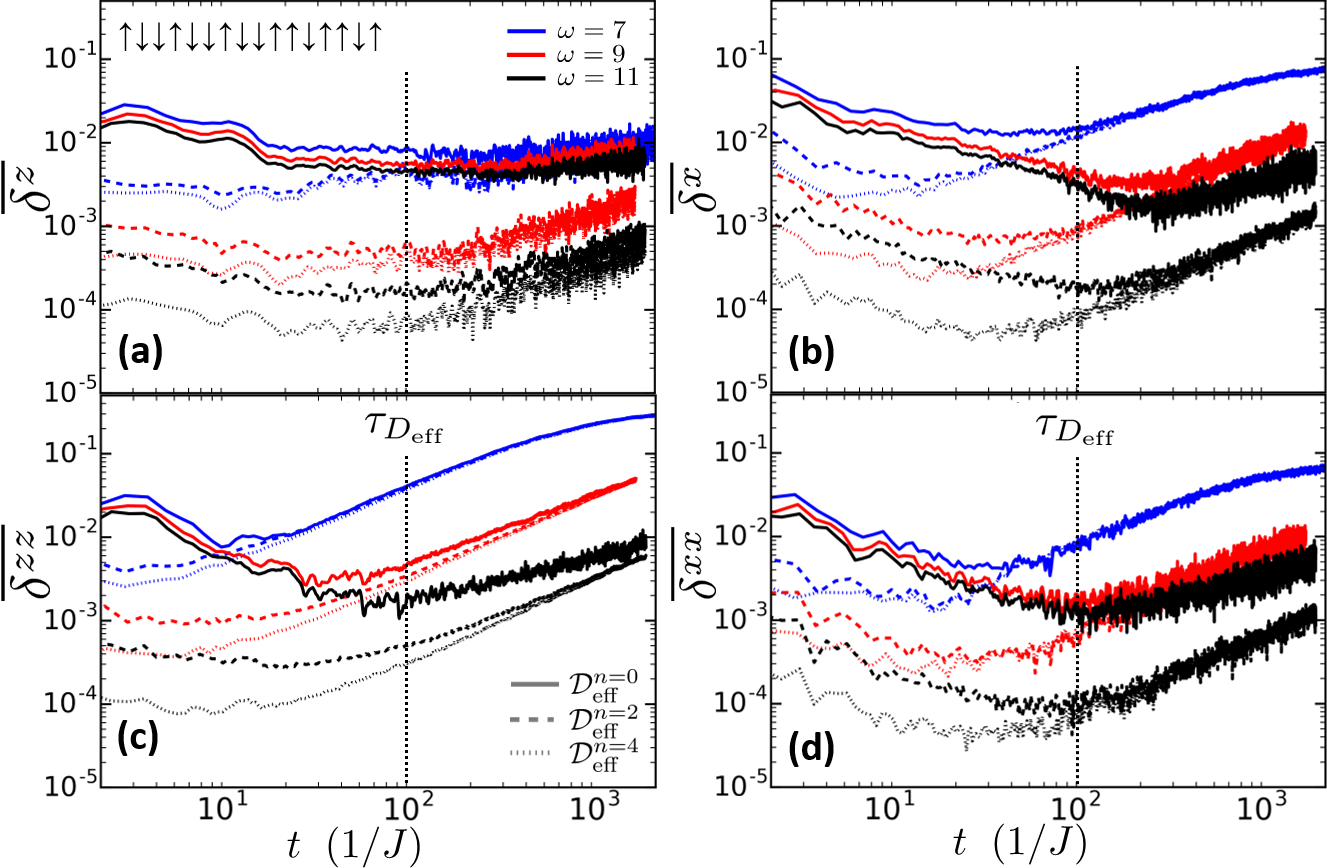}
    \caption{Difference in local operators, ($\sigma_i^z$ [a],$\sigma_i^x$ [b],$\sigma_i^z\sigma_{i+1}^z$ [c],$\sigma_i^x\sigma_{i+1}^x$ [d])  when evolved under the full evolution and $D_\eff^{n}$ of the long-range model. We observe a qualitatively similar picture to the results presented in \rfig{LocalShort}. This provides further evidence of $D_\eff$ being the approximate generator of stroboscopic time evolution. Similarly to \rfig{LocalShort}, the error dynamics becomes simpler for $t>\tau_{D_\eff}$, as extracted from \rfig{Plateau}.}
  \label{LocalLong}
\end{figure}

We now demonstrate that the results presented in \reffig{4} and \rfig{fig4SM} extend to other local operators of the system. In particular we will consider the operators $\sigma_i^z$, $\sigma_i^x$, $\sigma_i^z\sigma_{i+1}^z$ and $\sigma_i^x\sigma_{i+1}^x$ at site $i$, by measuring the errors $\delta^z_i$,$\delta^x_i$, $\delta^{zz}_i$ and $\delta^{xx}_i$ respectively, defined between time evolution under $\mathcal{D}_\eff^n$ and the full Hamiltonian. We then define $\overline{\delta^z}$,$\overline{\delta^x}$, $\overline{\delta^{zz}}$ and $\overline{\delta^{xx}}$ as the average of the errors over the all sites of the chain.
In analogy to \reffig{4} and \rfig{fig4SM}, we observe the emergence of a late time linear regime for all the considered operators, as shown in \rfig{LocalShort} (short-range) and \rfig{LocalLong} (long-range). The rate of growth of the linear regime decreases both with  increasing $n$ and $\omega$, consistent with $D_\eff$ being the approximate generator of time evolution. However, unlike the case when $\mathcal{O} = D_\eff^{(0)}/L$, the early time behavior has a  more complex structure. In these cases, we do not expect these local operators to be approximately conserved, so we observe  different, operator dependent, thermalization dynamics. 

\begin{figure}[H]
  \centering
  \includegraphics[width = 4.0in]{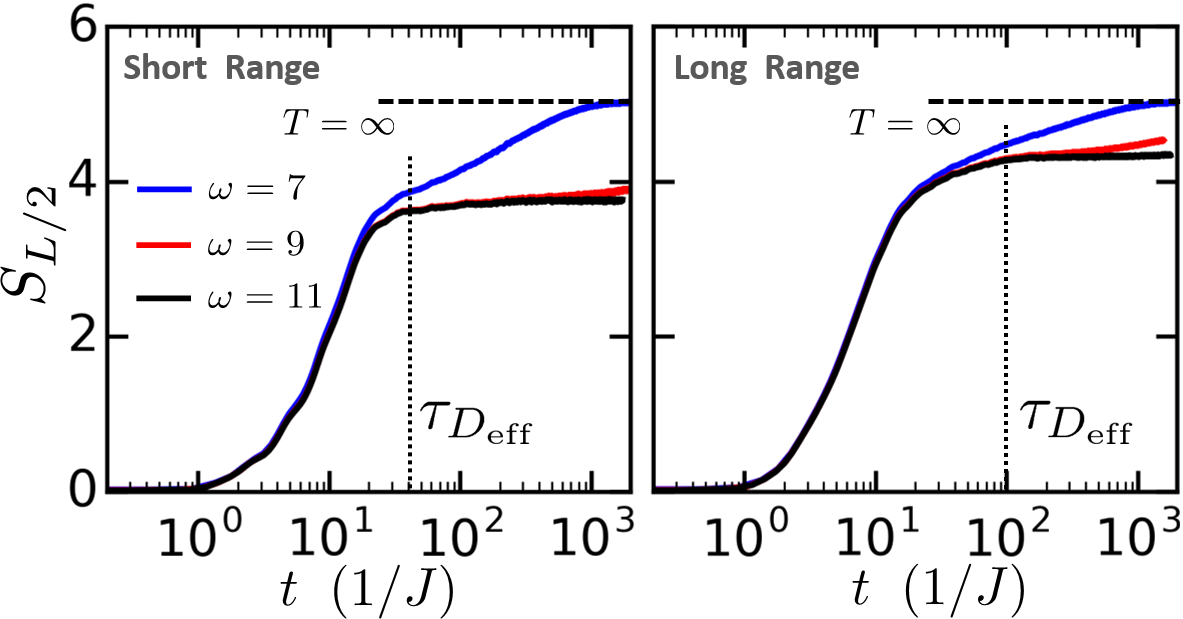}
    \caption{{\bf a[b])} Entanglement Entropy evolution for the full Hamiltonian for short-[long-] model. The thermalization of the system with respect to $D_\eff$ leads to the emergence of a plateau in the entanglement entropy. We define the time at which this plateau begins as $\tau_{D_\eff}$. We estimate this time scale for short- and long-range models as $\tau_{D_\eff}^s = 40$ and $\tau_{D_\eff}^\ell = 100$ respectively.}
  \label{Plateau}
\end{figure}

To corroborate that the early time behavior is due to  differences in short-time thermalization dynamics, we estimate $\tau_{D_\eff}$ as the time, the system approaches $S_{L/2}^P$. After this time, we observe that most of the error is given by an initial plateau followed by a linear  regime. For $t>\tau_{D_\eff}$, extracted in \rfig{Plateau}, the system has thermalized to $D_\eff$, so the error is dominated initially by the difference the thermal expectation of $\mathcal{O}$ with respect to $D_\eff$ or $\mathcal{D}_\eff^{n}$, until the linear growth in error from the difference between $D_\eff$ and the full evolution dominates.

Finally, we emphasize that the agreement we observe for the long-range model in \rfig{LocalLong} between the evolution under the full Hamiltonian and the different orders of $D_\eff$ at different frequencies provides further evidence of the existence of a prethermal effective Hamiltonian given by $D_\eff$ that approximately describes the time evolution of our system, even though no formal proofs exist for power-law interactions at present.

\begin{figure}[H]
  \centering
  \includegraphics[width = 3.7in]{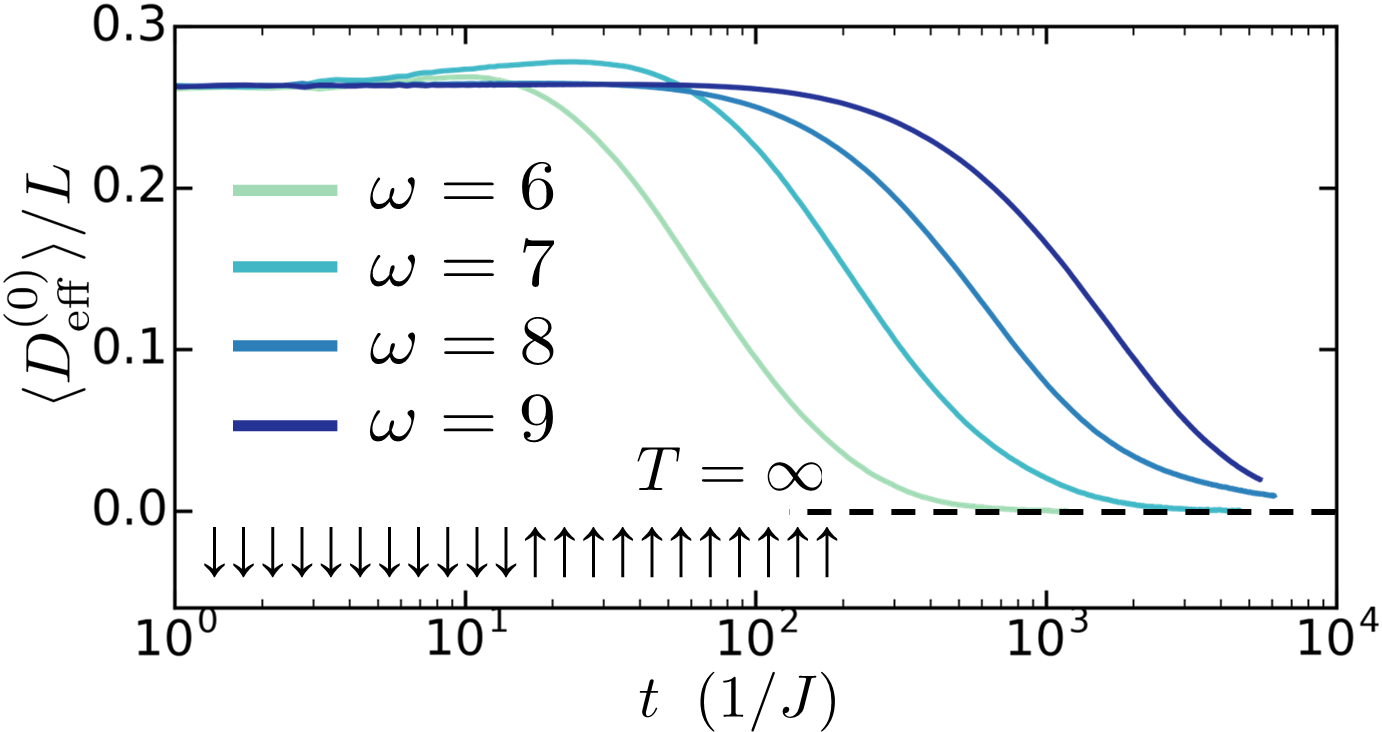}
  \caption{Exponentially long thermalization dynamics for  $L=22$ spins using an initial state on the ferromagnetic side of the spectrum. The process of heating toward the infinite temperature thermal ensemble actually shows up as a decrease of the energy density toward zero.}
  \label{fig:L22}
\end{figure}

\noindent For the data presented in Figure 2 of the main text, we focused on antiferromagnetic product states, where heating manifests as an increase in the energy density toward its infinite temperature value (namely, zero). In the case of ferromagnetic states, the process of heating actually corresponds to a decrease in the value of $\langle D_\eff^{(0)} \rangle /L$. As shown in \rfig{fig:L22} for a chain of $L=22$, one observes an exponential enhancement of the thermalization time from an initial state on the ferromagnetic side of the spectrum. 

\bibliography{pretherm}